%% file: DVpaper.tex
\documentclass[12pt]{JHEP3}
\usepackage[dvips]{graphics}

\def\be{\begin{equation}}
\def\ee{\end{equation}}

\def\lbldef#1#2{\expandafter\gdef\csname #1\endcsname {#2}}

\newcommand{\ev}[1]{\ensuremath{\left\langle #1 \right\rangle}}
\newcommand{\tr}{\ensuremath{\mathrm{tr}}}

\title{Gravitational F-terms of  ${\cal N}=1$ Supersymmetric
Gauge Theories}

\author{Ido Adam\footnote{e-mail: {\tt adamido@post.tau.ac.il}}\, and
Yaron Oz\footnote{e-mail: {\tt yaronoz@post.tau.ac.il}} \\
Raymond and Beverly Sackler Faculty of Exact Sciences \\
School of Physics and Astronomy \\
Tel-Aviv University , Ramat-Aviv 69978, Israel}

\abstract{We consider four-dimensional ${\cal N}=1$ supersymmetric
gauge theories in a supergravity background.  We use generalized
Konishi anomaly equations and R-symmetry anomaly to compute the exact
perturbative and non-perturbative gravitational F-terms.  We study two
types of theories: The first model breaks supersymmetry dynamically,
and the second is based on a $G_2$ gauge group.  The results are
compared with the corresponding vector models.  We discuss the
diagrammatic expansion of the $G_2$ theory.}

\preprint{hep-th/0407200 \\ TAUP-2776-04}

\begin{document}

\section{Introduction}

F-terms of four-dimensional supersymmetric
gauge theories in a supergravity and graviphoton backgrounds
have attracted much attention in recent years.
On the one hand they are related to certain exactly computable 
amplitudes of two gravitons and graviphotons.
On the other hand they are computed by second quantized partition functions
of topological strings  \cite{Antoniadis:1993ze},
and have an interesting mathematical structure \cite{Bershadsky:1993cx}.
Gravitational F-terms are directly related to the  partition function of
two-dimensional non-critical strings \cite{Dijkgraaf:2003xk,Ita:2004yn}.
Recently, gravitational F-terms have been related to the computation
of certain ${\cal N}=2$ black holes partition function \cite{Ooguri:2004zv}.

In this paper  
we will consider the gravitational F-terms
in the context of 
four-dimensional ${\cal N}=1$ supersymmetric gauge theories.
Dijkgraaf and Vafa suggested a matrix model description, where the gravitational
F-terms can be computed by summing up the 
non-planar matrix 
diagrams \cite{Dijkgraaf:2002dh}.
The assumption made is that the relevant fields are the glueball superfields $S_i$ and the
F-terms are holomorphic couplings of the glueball superfields to gravity.
The DV matrix proposal has been proven diagrammatically
in \cite{Ooguri:2003qp,Ooguri:2003tt}.

In this paper we will consider the 
gravitational F-terms of the form
\begin{eqnarray}
\Gamma_G & = & 
\int d^4x d^2\theta G_{\alpha \beta \gamma}
G^{\alpha \beta \gamma} 
 F_1 (S_i) \ ,
\label{Gammag2}
\end{eqnarray}
where $G_{\alpha\beta\gamma}$ is the  ${\cal N}=1$ 
Weyl superfield.
According to the DV proposal, $F_1 (S_i)$ is the partition function 
of the corresponding matrix model evaluated by
summing the genus one diagrams with $S_i$ being the 't Hooft
parameter.

The approach we will take is to
use generalized Konishi anomaly equations and R-symmetry anomaly
to compute
the exact perturbative and non-perturbative gravitational F-terms. We will consider
a vanishing graviphoton background.
In general, it is not clear in which cases the  generalized Konishi anomaly equations
are sufficient in order to determine the  gravitational F-terms.
We will study two
types of theories:
The first model breaks supersymmetry dynamically, and the second
is based on a gauge group that does not have a large $N_c$ expansion. We will consider 
a
$G_2$ gauge group.

In a model that breaks supersymmetry 
the chiral ring relations
cannot be used. This will be analysed following \cite{Brandhuber:2003va}, by
adding a certain deformation to the tree-level superpotential.
The model based on the $G_2$ gauge group does not have a large $N_c$ expansion.
This complicates the relation between the matrix (vector)
model computations and the gauge theory ones.
In both cases,  
we will compare the results 
to their counterparts in the corresponding vector models.
We will also discuss the diagrammatic expansion of the $G_2$ theory.

The paper is organized as follows. In section
\ref{sec:computational-scheme} the computational scheme for computing
the gravitational F-terms is reviewed following \cite{David:2003ke,Ita:2003kk}.
In section \ref{sec:DSB} the
gravitational F-term for the model that breaks supersymmetry dynamically  is computed, 
and compared with the corresponding vector model.
In section \ref{sec:G2SYM} the same computation and comparison are
performed for the $G_2$ SYM theory.
Details of the computations are presented for the two models in appendices A and
B respectively.
In appendix C we discuss the diagrammatics of the $G_2$ model. 

Other recent works on the computation of gravitational F-terms 
are \cite{Klemm:2002pa,Dijkgraaf:2002yn,David:2003ke,Alday:2003ms, Alday:2003dk,Gripaios:2003gw,Fuji:2004vf}.

\section{The Computational Scheme} \label{sec:computational-scheme}

In this section we will review the computational scheme for computing
the gravitational F-terms.

\subsection{Deformed Chiral Ring}

Consider first an  ${\cal N}=1$ supersymmetric 
gauge theory in flat space with a gauge group $G$ and some matter supermultiplets.
We will denote the four-dimensional Weyl spinor supersymmetry generators by 
$Q_{\alpha}$ and $\bar{Q}_{\dot{\alpha}}$.
Chiral operators are operators annihilated by
$\bar{Q}_{\dot{\alpha}}$.
For instance, the lowest component $\phi$ of a chiral superfield
$\Phi$ is a chiral operator. 
The OPE of two chiral operators is nonsingular and allows for the definition 
of the product of two chiral operators. The product of chiral operators is 
also a chiral operator. 
Furthermore, one can define a ring structure on the set of 
equivalence classes of chiral operators modulo operators of the form
$\{\bar{Q}_{\dot{\alpha}}$,$\cdots \left.\right]$.

Denote by $V$ the vector superfield in the adjoint representation
of $G$, by $\Phi$ chiral superfields in a representation 
$r$ of $G$ and by $\phi$ their lowest component. 
The field strength (spinor) superfield is $W_\alpha =
-\frac{1}{4}\bar D^2e^{-V}D_\alpha e^V$ and  is a chiral superfield.
One has
\begin{eqnarray}
\{W_\alpha^{(r)},W_\beta^{(r)}\}= 0\, , \quad W_\alpha^{(r)}\phi^{(r)} = 0
\ ,
\label{rel0} 
\end{eqnarray}
modulo $\{\bar{Q}_{\dot{\alpha}}$,$\cdots \left.\right]$ terms, 
where we noted that $\phi$ transforms in a 
representation $r$ of the
gauge group $G$, such that $W_\alpha^{(r)}= W_{\alpha}^aT^a(r)$ with $T^a(r)$
being the generators of the gauge group $G$ in the representation $r$.

Consider next the coupling of the supersymmetric gauge theory to 
a background ${\cal N}=1$
supergravity.
We denote by $G^{\alpha\beta\gamma}$ the  ${\cal N}=1$ 
Weyl superfield.
In the following we will denote by $W_\alpha$ 
the supersymmetric gauge field strength as well as its lowest component, 
the gaugino, and similarly for $G^{\alpha\beta\gamma}$.
The chiral ring relations (\ref{rel0}) are deformed to
\begin{eqnarray}
\{W_\alpha^{(r)},W_\beta^{(r)}\}=2G_{\alpha\beta\gamma}W^{\gamma{(r)}}\, , 
\quad W_\alpha^{(r)}\phi^{(r)}=0 \ .
\label{rel1} 
\end{eqnarray}
Together with  Bianchi identities of $N=1$ supergravity these relations generate 
all the relations in the deformed chiral ring.
Some 
relations that will be used later are \cite{David:2003ke}
\begin{eqnarray}
\label{ringrelgrav}
 && \left[W^2,W_\alpha \right]=0,\quad W^2W_\alpha
=-\frac{1}{3}G^2 W_\alpha,\quad W^2W^2=-\frac{1}{3}G^2 W^2 \ ,\\
 && G^4=(G^2)^2=0 \ . \nonumber
\end{eqnarray}
Throughout the paper we will follow the conventions used in \cite{Ita:2003kk}.

In addition to the above kinematical relations, one has
kinematical relations for the matter fields and dynamical relations 
from the variation of the tree level superpotential $W_{tree}$
\begin{eqnarray}
  \phi\frac{\partial W_{tree}}{\partial \phi}=0 \ .
\label{chirrelc}
\end{eqnarray}

\subsection{Konishi Anomaly Relations}

The classical chiral ring relations are, in general, modified
quantum mechanically.
The
classical relations arising 
from (\ref{chirrelc}) have a natural 
generalization, as anomalous Ward identities of the quantized matter 
sector in a classical gauge(ino) and supergravity background.
The classical Konishi equation reads
\begin{eqnarray}
\bar D^2J=\phi'\frac{\partial W_{tree}}{\partial \phi} \ ,
\end{eqnarray}
where $J$ is the generalized Konishi current and $\delta \phi =
\phi'(\phi)$
is the  generalized Konishi transformation.
This relation gets an anomalous contribution in the quantum
theory. It
takes the form 
\cite{Cachazo:2002ry,Konishi:1984hf,Konishi:1985tu}
\begin{eqnarray}
  \label{genkonishi}
\bar D^2J=
\phi'_i\frac{\partial W_{tree}}{\partial \phi_i}+\frac{1}{32\pi^2}\left(W_\alpha{}_i{}^j
W^\alpha{}_j{}^k+\frac{1}{3}G^2\delta^k_i\right)\frac{\partial\phi'_k}{\partial\phi_i}
\ ,
\end{eqnarray}
where $i$, $j$ and $k$ are gauge indices and their contraction
is in the appropriate representation. 

Since the divergence $\bar D^2J$ is $\bar{Q}$-exact it vanishes in a 
supersymmetric vacuum. Taking the expectation value of 
(\ref{genkonishi}) in a slowly varying gaugino background $S$,  we get 
the Konishi relations in a supergravity 
background given by $G^2$ 
\begin{eqnarray}
 \label{chirreld}
\left< \phi'_i\frac{\partial W_{tree}}{\partial \phi}\right>_S
+\left<\left(\frac{1}{32\pi^2}W_\alpha{}_i{}^j
W^\alpha{}_j{}^k+\frac{1}{32 \pi^2}\frac{G^2}{3}
\delta^k_i\right)\frac{\partial\phi'_k}{\partial\phi_i}\right>_S
= 0 \ . 
\end{eqnarray}
We will use this relation to determine the supergravity 
corrections to the chiral correlators, which in turn can be 
integrated to give the perturbative part of the gravitational
F-terms of the corresponding ${\cal N}=1$ gauge theory. Henceforth,
we absorb the factor of $\frac{1}{32 \pi^2}$ within $G^2$.

\subsection{Computation of Gravitational F-terms}

We are interested in the low energy description 
of a four-dimensional  ${\cal N}=1$ supersymmetric gauge theory in the
background of  ${\cal N}=1$ supergravity. 
The assumption is that the relevant field is the glueball superfield $S$ and the
F-terms are holomorphic couplings of the glueball superfield to gravity.

In the absence of supergravity, the only relevant F-term is the effective glueball
superpotential 
\begin{equation}
\Gamma_0=\int d^4x d^2 \theta W_{eff}(S) \ ,
\end{equation}
where 
\begin{equation}
S=-\frac{1}{32 \pi^2}Tr\, W_{\alpha}W^{\alpha} \ .
\end{equation}
In the matrix model description $\Gamma_0$ is computed by summing up planar diagrams and 
adding a non-perturbative   Veneziano-Yankielowicz superpotential \cite{Veneziano:1982ah}.

When coupled to supergravity
there is a  gravitational F-term
of the form 
\begin{eqnarray}
 \label{eq:GC1} 
  \Gamma_1=\int d^4x d^2 \theta W_1(S) G^2 \ .
\end{eqnarray}
In the matrix model description it is computed by summing up non-planar diagrams and 
adding a non-perturbative contribution. 
Note that terms with higher powers of $G$ vanish due to the chiral ring relation 
(\ref{ringrelgrav}).

\subsubsection*{Computation of $W_1(S)$}
\label{pertpartcorr}

Consider the supersymmetric gauge theory with  a tree level superpotential
\begin{eqnarray}
\label{treelevel}
W_{tree}=\sum_I g_I \sigma_I \ ,
\end{eqnarray}
where  $\sigma_I$ are gauge invariant chiral operators and $g_I$ the
tree level couplings.
The gradient equations for the 
holomorphic part of the effective action read
\begin{eqnarray}
  \label{gradeqn}
\frac{\partial \left(W_{eff}+G^2 W_1 \right)}{\partial g_I}=\left< \sigma_I\right>_S  \ .
\end{eqnarray}
The expectation values are taken in a slowly varying (classical)
gaugino and gravitino background.

As first discussed in \cite{Cachazo:2002ry}, for a gauge theory in 
the absence of a supergravity background the Konishi relations (\ref{chirreld})
can be used to solve for the expectation values $\left< \sigma_I\right>_S$
as a function of $S$ and the tree level couplings. One can then integrate
(\ref{gradeqn}) to determine the dependence of $W_{eff}$ on the tree level couplings.
For the gravitational coupling $W_1(S)$ a similar reasoning applies. However,
we will have to take into account the effects of the
supergravity background on the correlators of chiral operators.

In the absence of gravity the correlators of chiral operators
factorize
\begin{eqnarray}
  \label{eq:mesoncorr}
  \left< \sigma_I\sigma_J \right>&=&\left< \sigma_I \right>\left<
  \sigma_J\right> \ .
\end{eqnarray}
In a matrix model description, corresponding to gauge theories
with large $N_c$ expansion, this is the feature of the planar limit.
Here and in some of the equations in the following 
we omit for simplicity the subscript $S$.
Eq.~(\ref{eq:mesoncorr}) can be used in the relations (\ref{chirreld}) in order to solve 
for $\left< \sigma_I\right>_S$ as a function of $(S,g_I)$.

However, in the presence of supergravity 
the chiral correlators do not factorize, and instead we have 
\begin{eqnarray}
  \label{eq:mesoncorr2}
  \left< \sigma_I\sigma_J \right>&=&\left< \sigma_I\sigma_J \right>_c
                      +\left< \sigma_I \right>\left< \sigma_J\right>
                      \ ,  
\end{eqnarray}
with analogous relations for correlators with more chiral operators.
Also, the one point functions have to be expanded in $G^2$ as  
\begin{eqnarray}
  \label{eq:epans}
  \left< \sigma_I\right>&=& \left<\sigma_I\right>_1+ G^2\left<
  \sigma_I\right>_2 \ .
\end{eqnarray}
Note that this expansion is exact in the chiral ring due to the fact that 
$G^4$ vanishes modulo ${\bar D}$ exact terms. 
Thus, we have to express $\left< \sigma_I\right>_1$ and  $\left< \sigma_I\right>_2$ 
as functions of $S$ and $g_I$.

In the next section we will show explicitly that there are enough  
relations (\ref{chirreld}) to solve for $\left< \sigma_I
\right>,\left< \sigma_J\right>$ as well as 
for the connected correlators $\left< \sigma_I\sigma_J \right>_c$.
The perturbative part of the gravitational coupling $W_1(S)$ is then obtained by
integrating the gravitational contribution $\left< \sigma_K\right>_2$
in (\ref{eq:epans})
for the $\left< \sigma_K\right>$ appearing in the tree level potential
(\ref{treelevel}), with respect to the couplings $g_K$.

Note, that a crucial  ingredient in the analysis is the assumption that 
connected correlators of three or more chiral operators vanish in the 
gravitationally deformed chiral ring.

The procedure outlined above determines $W_1(S)$ up to an integration constant 
independent of the couplings $g_I$. The
integration constant can be determined by the one loop
exact $U(1)_R$ anomaly, as will be done later.

\section{Dynamical Supersymmetry Breaking} \label{sec:DSB}
The model considered is an $\mathcal{N}=1$ SYM theory with an
$Sp(N_c)$ gauge group coupled to $2N_f=2(N_c+1)$ fundamental chiral
multiplets $Q_a^i$ ($a=1,\dots,2N_c$ is the gauge index and the flavor
index is $i=1,\dots,2N_f$) and a chiral gauge singlet $S_{ij}$
antisymmetric in the flavor indices
\cite{Izawa:1996pk,Intriligator:1996pu}.  The gauge invariant matter
in the theory is the $S_{ij}$ and the mesons $M^{ij}=Q^{ai} Q_{aj}$,
which are antisymmetric in the flavor indices $i$ and $j$.

The tree-level superpotential is taken to be
\begin{equation}
W_\mathrm{tree} = \lambda S_{ij} M^{ij} - m J^{ij} S_{ij} \;,
\end{equation}
where $J=\mathbf{1}_{N_c} \otimes i \sigma^2$ is the symplectic form.

This theory has no supersymmetric vacuum, so the chiral ring relations
cannot be used. This can be remedied following \cite{Brandhuber:2003va}, by
adding a deformation to the tree-level superpotential giving mass to
$S_{ij}$:
\begin{equation} \label{eq:DSBsuperpotential}
W_\mathrm{tree} = \lambda S_{ij} M^{ij} - m J^{ij} S_{ij}
+ \alpha S_{ij} S^{ij} \;,
\end{equation}
where $S^{ij}=S_{kl} J^{ki} J^{lj}$. This deformation adds a
supersymmetric vacuum and enables the use of the chiral ring.

\subsection{Calculation of $W_1$}
\subsubsection*{The Perturbative Superpotential}
Using Konishi transformations either in $Q_a^i$ or $S_{ij}$ (the
detailed computation is in appendix \ref{sec:details-DSB-model}) the
following Konishi anomaly equations are obtained
\begin{eqnarray}
S \delta_i^l& = & 2 \lambda \ev{S_{ij} M^{lj}} + \frac{2 N_c}{3} G^2
\delta_i^l \;, \label{eq:DSBKonishi1} \\
0 & = & \lambda \ev{S_{lm} M^{ij}} - m \ev{S_{lm}} J^{ij}
+ 2 \alpha \ev{S_{lm} S^{ij}} + \frac{1}{6} G^2 (\delta^i_l
\delta^j_m - \delta^i_m \delta^j_l) \;, \label{eq:DSBKonsihi2}
\\
S \ev{M^n} \delta^l_i & = & 2 \lambda \ev{S_{ij} M^{lj} M^n} + \frac{2 N_c}{3} G^2 \ev{M^n}
\delta^l_i + \frac{2 n}{3} G^2 (J^{-1})_{ij} \ev{M^{lj} M^{n-1}} \;,
\label{eq:DSBKonishi3} \\
0 & = & \lambda \ev{S_{ij} M^{lm} \tilde S^n} - m J^{lm} \ev{S_{ij} \tilde
S^n} + 2 \alpha \ev{S_{ij} S^{lm} \tilde S^n} + \nonumber \\
&& {} + \frac{1}{6} G^2 \ev{\tilde S^n} (\delta^l_i \delta^m_j
- \delta^l_j \delta^m_i) +
\frac{n}{3} G^2 J^{lm} \ev{S_{ij} \tilde S^{n-1}} \;,
\label{eq:DSBKonishi4} \\
S \ev{\tilde S^n} \delta^l_i & = & 2 \lambda \ev{S_{ij} M^{lj} \tilde
S^n} + \frac{2 N_c}{3} G^2 \ev{\tilde S^n} \delta^l_i \;,
\label{eq:DSBKonishi5} \\
0 & = & \lambda \ev{S_{ij} M^{lm} M^n} - m J^{lm} \ev{S_{ij} M^n}
+ 2 \alpha \ev{S_{ij} S^{lm} M^n} \nonumber \\
&& {} + \frac{1}{6} G^2 \ev{M^n}
(\delta^l_i \delta^m_j - \delta^l_j \delta^m_i) \;,
\label{eq:DSBKonishi6} \\
S \ev{S_{ki}} & = & 2 \lambda \ev{S_{kl} M^{lm} S_{im}} + \frac{2
N_c}{3} G^2 \ev{S_{ki}} \;, \label{eq:DSBKonishi7}
\end{eqnarray}
where $\tilde S = J^{ij} S_{ij}$ and $M=J^{ij} M_{ij}$.

Assuming flavor symmetry and that all the connected
three-point-functions vanish, these equations can be solved for the
correlation functions. Picking the solution corresponding to the
massive vacuum, in which the chiral multiplets are massive, the order
$G^2$ terms of the relevant connected correlation functions (denoted
by $\ev{\dots}^g$) are
\begin{eqnarray}
\ev{S_{ij} M^{ij}}_c^g & = & \frac{N_f (2 N_f - 1) \left( -m^2 + 4
\alpha S + m \sqrt{m^2 - 4 \alpha S} \right)}{6 \lambda \left( m^2 - 4
\alpha S \right)} \;, \\
\ev{S_{ij}}^g & = & \frac{-4 N_c \sqrt{m^2 - 4 \alpha
S} + (2 N_f - 1) \left( -m + \sqrt{m^2 - 4 \alpha S}\right)}{12 \left(
m^2 - 4 \alpha S \right)} (J^{-1})_{ij} \;, \\
\ev{M^{ij}}^g & = & \frac{\alpha \left[ -4 N_c \sqrt{m^2 -4 \alpha S} + (2 N_f
- 1) (-m + \sqrt{m^2 -4 \alpha S}) \right]}{6 \lambda (m^2 -4 \alpha
S)} J^{ij} \;, \\
\ev{S_{ij} S^{ij}}_c^g & = & -\frac{2 N_f (2 N_f - 1) \left( m^2 - 4
\alpha S + m \sqrt{m^2 -4 \alpha S} \right)}{24 \alpha \left( m^2 - 4
\alpha S \right)} \;.
\end{eqnarray}

The gradient equations (\ref{gradeqn}) in this model read
\begin{eqnarray}
\frac{\partial W_1}{\partial \lambda} & = & \ev{S_{ij} M^{ij}}^g \;,
\\
\frac{\partial W_1}{\partial m} & = & -J^{ij} \ev{S_{ij}}^g \;, \\
\frac{\partial W_1}{\partial \alpha} & = & \ev{S_{ij} S^{ij}}^g \;.
\end{eqnarray}
These can be integrated in order to obtain the gravitational F-term up
to a function of $S$, which is independent of the couplings. For
$N_f=N_c+1$ --- the case of unbroken supersymmetry --- one obtains
\newpage
\begin{eqnarray}
W_1 & = & \frac{N_f}{6} \Bigg[ (2 N_f - 3) \log \left( -\frac{4
\alpha}{m^2} S \right) - (2 N_f -3) \log \left( 1 + \sqrt{1 - \frac{4
\alpha}{m^2} S} \right) \nonumber \\
&& {} - (2 N_f - 1) \log \sqrt{1 - \frac{4 \alpha}{m^2} S} - 2 \log m
- 4 N_c \log \lambda \Bigg] + C(S) \;.
\end{eqnarray}

\subsubsection*{The Non-perturbative Contribution}
Using the appendix of \cite{Anselmi:1998am}, the R-symmetry anomaly for
this model is given by
\begin{equation}
\mathcal{A} = -\frac{1}{3} \left[ N_c \left( 2 N_c + 1 \right) -
\frac{4}{3} N_f N_c - \frac{1}{3} N_f \left( 2 N_f - 1 \right) \right] G^2
\end{equation}
in the convention $\frac{1}{32 \pi^2} G^2 \to G^2$.

The term in the action that reproduces this anomaly is similar to the
one in \cite{Ita:2003kk}:
\begin{eqnarray}
\Gamma_1 (S, G^2) & = & \int d^4 x d^2 \theta \frac{1}{6} \bigg[ N_c
\left( 2 N_c + 1 \right) \log \frac{S}{\Lambda_1^3} + \frac{4}{3} N_f
N_c \log \frac{\Lambda_1}{\mu} \nonumber \\
&& {} + \frac{1}{3} N_f (2 N_f - 1 ) \log \frac{\Lambda_1}{\alpha}
\bigg] G^2 \;,
\end{eqnarray}
where $\Lambda_1$ is the supergravity scale and $\mu$ is the mass
scale for the massless matter multiplets in the fundamental representation. The
R-symmetry transformation is taken to be
\begin{eqnarray*}
\theta' = e^{-i \alpha} \theta \;, &&  \bar \theta' = e^{i \alpha}
\bar \theta \;, \\ 
S'(x, \theta', \bar \theta') & = & e^{-2 i \alpha} S(x, \theta, \bar
\theta) \;, \\
G'^2 (x, \theta', \bar \theta')
& = & e^{-2 i \alpha} G^2 (x, \theta, \bar \theta) \;.
\end{eqnarray*}

In the massive vacuum, in which the massive $S_{ij}$ obtains an
expectation value and endows the matter in the fundamental with a mass
through its quadratic term in the superpotential, the massive matter
multiplets decouple from the gauge sector in the IR, and only the
gauge part of the anomaly has to be matched in the perturbative
superpotential. After matching, the order $G^2$ F-term is
\begin{eqnarray}
W_1 & = & \frac{N_f}{6} \Bigg[ (2 N_f -3 ) \log \left( -\frac{4
\alpha}{m^2} S \right)  - (2 N_f - 3) \log \left( 1 + \sqrt{1 -
\frac{4 \alpha}{m^2} S} \right) \nonumber \\
&& {} - (2 N_f - 1) \log \sqrt{1 - \frac{4 \alpha}{m^2} S} -
2 \log \frac{m}{\Lambda_1} - 4 N_c \log \lambda \Bigg] +
\frac{1}{6} \log \frac{S}{\Lambda_1^3} \;.
\end{eqnarray}

\subsection{The Vector Model Ward Identities}
The partition function of the vector model corresponding to this model
is
\begin{equation}
Z = \int [dQ^a_i] [dS_{ij}] e^{-\frac{1}{g} W_\mathrm{tree} (Q,
S_{ij})} \;,
\end{equation}
where the action $W_\mathrm{tree}$ is given by
(\ref{eq:DSBsuperpotential}). The coupling constant $g$ is introduced
for relating the vector model Ward identities with the gauge theory
Konishi anomaly equations.

A transformation of the form $Q_a^i \to Q_a^i + \delta Q_a^i$ with
$S_{ij}$ unchanged generates the Ward identity
\begin{equation}
\ev{\frac{\partial W_\mathrm{tree}}{\partial Q_a^i} \delta Q_a^i} = g
\ev{\frac{\partial \delta Q_a^i}{\partial Q_a^i}} \;.
\end{equation}
Similarly, a transformation of the form $S_{ij} \to S_{ij} + \delta
S_{ij}$ with $Q_a^i$ fixed yields the Ward identity
\begin{equation}
\ev{\frac{\partial W_\mathrm{tree}}{\partial S_{ij}} \delta S_{ij}} =
g \ev{\frac{\partial \delta S_{ij}}{\partial S_{ij}}} \;.
\end{equation}

Using the same transformations leading to the gauge theory anomaly
equations (\ref{eq:DSBKonishi1})--(\ref{eq:DSBKonishi7}) we obtain
the following Ward identities
\begin{eqnarray}
&& 2 \lambda \ev{S_{ij} M^{lj}} = 2 g N_c \delta^l_i \;, \\
&& \lambda \ev{S_{lm} M^{ij}} - m J^{ij} \ev{S_{lm}} + 2 \alpha
\ev{S_{lm} S^{ij}} = \frac{g}{2} \left( \delta^i_l \delta ^j_m -
\delta^i_m \delta^j_l \right) \;, \\
&& 2 \lambda \ev{S_{ij} M^{lj} M^n} = 2 g N_c \ev{M^n} \delta^l_i + 2
n g (J^{-1})_{ij} \ev{M^{lj} M^{n-1}} \;, \\
&& \lambda \ev{S_{lm} M^{ij} \tilde S^n} - m J^{ij} \ev{S_{lm} \tilde
S^n} + 2 \alpha \ev{S_{lm} S^{ij} \tilde S^n} = \frac{g}{2}
\ev{\tilde S^n}  \left( \delta^i_l \delta^j_m - \delta^i_m \delta^j_l
\right) \nonumber \\
&&{} + n g \ev{S_{lm} \tilde S^{n-1}} J^{ij} \;, \\
&& 2 \lambda \ev{S_{ij} M^{lj} \tilde S^n} = 2 g N_c \ev{\tilde S^n}
\delta^l_i \;, \\
&& \lambda \ev{S_{lm} M^{ij} M^n} - m J^{ij} \ev{S_{lm} M^n} + 2 \alpha
\ev{S_{lm} S^{ij} M^n} = \nonumber \\
&& \qquad \frac{g}{2} \ev{M^n} \left( \delta^i_l
\delta^j_m - \delta^i_m \delta^j_l \right) \,, \\
&& 2 \lambda \ev{S_{km} M^{ml} S_{il}} = 2 g N_c \ev{S_{ki}} \;.
\end{eqnarray}
Comparison of these equations with their gauge theory counterparts
(\ref{eq:DSBKonishi1})--(\ref{eq:DSBKonishi7}) yields that the
gravitational genus one F-term is related to the vector model free
energy by the relations
\begin{equation}
g = -\frac{1}{3} G^2 \;, \quad 2 g N_c = S - \frac{2 N_c}{3} G^2 \;.
\end{equation}

Hence, the contribution of planar diagrams to the perturbative part of
the genus one F-term is given by a shift $S \to S-\frac{2 N_c}{3} G^2$
\cite{David:2003ke}, in the perturbative part of the effective
superpotential of \cite{Brandhuber:2003va}, taken about the
massive vacuum
\begin{eqnarray}
W_1^\mathrm{planar} & = & -\frac{2 N_c}{3} \frac{\partial
W_\mathrm{eff}^\mathrm{pert}}{\partial S} = \frac{N_f N_c}{3} \bigg[
-1 + 2 \log \alpha - 2 \log \left( m + \sqrt{m^2 - 4 \alpha S} \right)
\nonumber \\
&& {} - 2 \log \lambda \bigg] \;,
\end{eqnarray}
where the number of colors $N_c$ has been specified explicitly.
The perturbative part of $W_1$ proportional to $N_c$ is 
\begin{displaymath}
\frac{2 N_f N_c}{3} \left[ \log \left( -4 \alpha S \right) - \log
\left( m + \sqrt{m^2 - 4 \alpha S} \right) - \log \lambda \right]
\end{displaymath}
and all of it except the $\frac{2 N_f N_c}{3} \log (-4 S)$ term is
accounted for by the planar contribution.

\section{$G_2$ SYM with Three Flavors} \label{sec:G2SYM}
This is an $\mathcal{N}=1$ SYM theory with the gauge group $G_2$ with
three flavors of chiral matter in the real fundamental $\mathbf{7}$
representation considered in \cite{Brandhuber:2003va}. The chiral superfields
are denoted by $Q^i_I$ (henceforth $i,j,k,\dots=1,\dots,7$ denote
color indices and $I,J,K,\dots=1,\dots,3$ are flavor indices).  The
gauge invariant operators of this theory are the six mesons
$X_{IJ}=\delta^{ij} Q^i_I Q^j_J$ and the single baryon $Z=\psi^{ijk}
\epsilon_{IJK} Q^i_I Q^j_J Q^k_K$, where $\psi^{ijk}$ is the $G_2$
invariant $3$-form.

The tree-level superpotential
\begin{equation} \label{eq:G2-tree-superpotential}
W_\mathrm{tree} = m^{IJ} X_{IJ} + \lambda Z
\end{equation}
is taken with the mass matrix $m^{IJ}=m \delta^{IJ}$, leaving the
flavor symmetry intact.

\subsection{Computation of $W_1$}
\subsubsection*{The Perturbative Superpotential}
The simplest Konishi equations for this theory (more explicit details are in
appendix \ref{sec:G_2details}) 
\begin{eqnarray}
2 S \delta_{IJ} & = & 2 m \ev{X_{IJ}} + \lambda \ev{Z} \delta_{IJ} +
\frac{7}{3} G^2 \delta_{IJ} \;, \label{eq:G2Konishi1} \\
0 & = & 2 m \ev{Z} + 6 \lambda \left( \ev{X_I^{\phantom{I}I}
X_J^{\phantom{J}J}} - \ev{X_{IJ} X^{IJ}} \right) \;,
\label{eq:G2Konishi2} \\
2 S \ev{\tr(X^n)} \delta_{IL} & = & 2 m \ev{X_{IL} \tr(X^n)} + \lambda \ev{Z \tr(X^n)} \delta_{IL} +
\frac{7}{3} G^2 \ev{\tr (X^n)} \delta_{IL}
\nonumber \\
&& {} + \frac{2n}{3} G^2 \ev{(X^n)_{LI}} \;, \label{eq:G2Konishi3} \\
0 & = & 2 m \ev{Z \tr (X^n)} + 6 \lambda \left(
\ev{X_I^{\phantom{I}I} X_J^{\phantom{J}J} \tr (X^n)} - \ev{X_{IJ}
X^{IJ} \tr (X^n)} \right) \nonumber \\
&& {} + \frac{2 n}{9} G^2 \ev{Z \tr (X^{n-1})} \;, \label{eq:G2Konishi4} \\
2 S \ev{X_J^{\phantom{J}I}} & = & 2 m \ev{X_{JK} X^{KI}} + \lambda
\ev{X_J^{\phantom{J}I} Z} + \frac{1}{3} G^2 \ev{X_K^{\phantom{K}K}}
\delta_J^{\phantom{J}I} \nonumber \\
&& {} + \frac{8}{3} G^2 \ev{X_J^{\phantom{J}I}} \;. \label{eq:G2Konishi5}
\end{eqnarray}
These equations can be solved and the needed correlation functions in
the Higgsed vacuum found in \cite{Brandhuber:2003va} are
\begin{eqnarray}
\ev{X_I{}^I}^g & = & -\frac{3 m^3 - 72 \lambda^2 S + 11 \sqrt{m^6 -
36 \lambda^2 m^3 S}}{4 m \left( m^3 - 36 \lambda^2 S \right)} \;, \\
\ev{Z}^g & = & -\frac{11 m^3 - 432 \lambda^2 S - 11 \sqrt{m^6 - 36
\lambda^2 m^3 S}}{6 \lambda \left( m^3 - 36 \lambda^2 S \right)} \;.
\end{eqnarray}

Utilizing the gradient equations (\ref{gradeqn}) in this case
\begin{eqnarray}
\frac{\partial W_1}{\partial m} & = & \ev{X_I^{\phantom{I}I}}^g \;, \\
\frac{\partial W_1}{\partial \lambda} & = & \ev{Z}^g \;,
\end{eqnarray}
the order $G^2$ correction to the superpotential can be integrated
in order to obtain
\begin{eqnarray}
W_1 & = & - \frac{1}{12} \left[ 22 \log \left( 1+ \sqrt{1 - \frac{36
\lambda^2}{m^3} S} \right) + \log \left( 1 - \frac{36 \lambda^2}{m^3}
S \right) + 42 \log m \right] \nonumber \\
&& {} + C_1 (S) \;.
\end{eqnarray}

\subsubsection*{The Non-perturbative Part}
The non-perturbative part of $W_1$ is found as in \cite{Ita:2003kk} by
requiring that in the limit $\lambda \to 0$
\begin{equation}
\Gamma_1 = \int d^4 x d^2 \theta W_1 (S) G^2
\end{equation}
reproduce the $U(1)_R$ anomaly.

The matter fields are all integrated out so we have to match only the
anomaly in the gauge sector. Comparing with (24) in \cite{Ita:2003kk}
and taking only terms of order $G^2$ we have in the gauge sector the
anomaly
\begin{equation}
\mathcal{A} = - \frac{1}{3} G^2 (\mathrm{rank}) = - \frac{14}{3} G^2 \;,
\end{equation}
since $G_2$ has $14$ generators.

The $U(1)_R$ transformation is defined by
\begin{eqnarray}
\theta' = e^{-i \alpha} \theta \;, &&
\bar \theta' = e^{i \alpha} \bar \theta \;, \nonumber \\
S'(x, \theta', \bar \theta') & = & e^{-2 i \alpha} S(x, \theta, \bar
\theta) \;, \\
G'^2 (x, \theta', \bar \theta') & = & e^{-2 i \alpha} G^2 (x, \theta, \bar
\theta) \;. \nonumber
\end{eqnarray}
The term $W_1^\mathrm{non-pert} G^2=\frac{14}{6} G^2 \log
\frac{S}{\Lambda_1^3}$ has the required anomaly so we take $C_1
(S)=\frac{7}{3} \log \frac{S}{\Lambda_1^3}$ and for solution
(\ref{eq:G2solution2}) the correction becomes
\begin{eqnarray}
W_1 (S) & = & - \frac{1}{12} \left[ 22 \log \left( 1 + \sqrt{1 -
\frac{36 \lambda^2}{m^3} S} \right) +
\log \left( 1 - \frac{36 \lambda^2}{m^3} S \right) \right. \nonumber\\
&& \left. \vphantom{\left( 1 - \sqrt{\frac{36 \lambda^2}{m^3} S} \right)} +
42 \log m - 28 \log \frac{S}{\Lambda_1^3} \right] \;.
\end{eqnarray}

In the $\lambda \to 0$ limit
\begin{equation}
\Gamma_1 \to \frac{1}{12} \int d^4 x d^2 \theta \left(
28 \log \frac{S}{\Lambda_1^3} - 42 \log m \right) G^2 \;,
\end{equation}
whose transformation is
\begin{equation}
\delta_{U(1)_R} \Gamma_1 \to - \frac{14}{3} i \alpha G^2\;.
\end{equation}

As argued in \cite{Ita:2003kk}, the same scale has to be used throughout
the $G^2$ term. Hence, dimensionality is taken care of in the
expression
\begin{eqnarray}
W_1 & = & -\frac{1}{12} \left[ 22 \log \left( 1 + \sqrt{1 -
\frac{36 \lambda^2}{m^3} S} \right) +
\log \left( 1 - \frac{36 \lambda^2}{m^3} S \right) \right. \nonumber \\
&& \left. \vphantom{\left( \sqrt{\frac{36 \lambda^2}{m^3} S} \right)}
+ 42\log \frac{m}{\Lambda_1} -28 \log \frac{S}{\Lambda_1^3} \right] \;.
\end{eqnarray}

\subsection{Comparison with the Vector Model}
The partition function of the corresponding vector model is
\begin{equation} \label{eq:G2partition-fn}
Z=\int dQ e^{-\frac{1}{g} W_\mathrm{tree} (Q)} \;,
\end{equation}
where the action is given by the tree-level superpotential of the
gauge theory (\ref{eq:G2-tree-superpotential}) and $g$ is a coupling
that should be replaced with a function of $S$ and $G^2$ in order to
reproduce the gauge theory anomaly equations.

In general, a transformation $Q^i_I \to Q^i_I+\delta Q^i_I$ generates
the vector model Ward identity
\begin{equation}
\ev{ \frac{\partial W_\mathrm{tree}}{\partial Q^i_I} \delta Q^i_I} = g
\ev{\tr \frac{\partial \delta Q^j_J}{\partial Q^i_I}} \;.
\end{equation}

The vector model Ward identity corresponding to the anomaly equation
(\ref{eq:G2Konishi1}) is
\begin{equation} \label{eq:G2Ward-identity1}
2 m \ev{X_{IJ}} + \lambda \ev{Z} \delta_{IJ} = \hat 7 g \delta_{IJ} \;,
\end{equation}
where $\hat 7$ denotes a factor of 7 coming from a trace on the
$\mathbf{7}$ representation of $G_2$.
The vector model counterpart of (\ref{eq:G2Konishi2}) is the Ward
identity
\begin{equation} \label{eq:G2Ward-identity2}
2 m \ev{Z} + 6 \lambda \left( \ev{X_I{}^I X_J{}^J} - \ev{X_{IJ} X^{IJ}}
\right) = 0 \;.
\end{equation}
This identity is actually identical to the anomaly equation.
The transformation leading to (\ref{eq:G2Konishi3}) yields the Ward
identity
\begin{equation} \label{eq:G2Ward-identity3}
2 m \ev{X_{IJ} \tr(X^n)} + \lambda \ev{Z \tr(X^n)} \delta_{IJ} =
\hat 7 g \ev{\tr(X^n)} \delta_{IJ} + 2 n g \ev{(X^n)_{IJ}} \;.
\end{equation}
Applying the same transformation as in (\ref{eq:G2Konishi4}) one
obtains the vector model equation
\begin{equation} \label{eq:G2Ward-identity4}
2 m \ev{Z \tr (X^n)} + 6 \lambda \left( \ev{X_I{}^I X_J{}^J \tr (X^n)}
- \ev{X_{IJ} X^{IJ} \tr (X^n)} \right) = \frac{2 n}{3} g \ev{Z \tr
(X^{n-1})} \;.
\end{equation} 
Finally, the analog of (\ref{eq:G2Konishi5}) is
\begin{equation} \label{eq:G2Ward-identity5}
2 m \ev{X_{IK} X^K{}_J} + \lambda \ev{X_{IJ} Z} = \hat 7 g \ev{X_{IJ}}
+ g \ev{X_{IJ}} + g \ev{X_K{}^K} \delta_{IJ} \;.
\end{equation}

Comparison of the Ward identities
(\ref{eq:G2Ward-identity1})--(\ref{eq:G2Ward-identity5}) with the
anomaly equation (\ref{eq:G2Konishi1})--(\ref{eq:G2Konishi5}) yields
the following identifications
\begin{equation} \label{eq:G2-g-S-relation}
g = -\frac{1}{3} G^2 \;, \quad \hat 7 g = 2 S - \frac{7}{3} G^2 \;.
\end{equation}

The gauge group $G_2$ does not admit large-$N_c$ expansion.
Thus, it is not clear how to directly compare the gauge theory F-terms computation and 
the vector model diagrammatic expansion.  
The relation between the gauge theory anomaly equations
and the vector model Ward identities
suggests, that a method of comparison should exist.
In particular, one may hope to identify which diagrams contribute to which
F-term. 
We have given some details of the $G_2$ diagrammatics in 
\ appendix \ref{sec:G2-diagrammatics}. 
So far, we have not found a direct comparison scheme.

\section*{Acknowledgments}

We would like to thank H. Ita and H. Nieder for valuable discussions.
This work is supported by the US-Israel BSF.

\appendix
\section{Details for the DSB model} \label{sec:details-DSB-model}
Using the Konishi transformation
$\delta Q_a^i=\epsilon^i_{\phantom{i}j} Q_a^j$, the following equation
is obtained in the chiral ring
\begin{equation}
2 \lambda \ev{S_{ij} M^{lj}} + \frac{2 N_c}{3} G^2 \delta_i^l = S
\delta_i^l \;.
\end{equation}
The transformation $\delta S_{ij} = \epsilon_{ij}^{\phantom{ij}lm} S_{lm}$
leads to the equation
\begin{equation}
\lambda \ev{S_{lm} M^{ij}} - m \ev{S_{lm}} J^{ij}
+ 2 \alpha \ev{S_{lm} S^{ij}} + \frac{1}{6} G^2 (\delta^i_l
\delta^j_m - \delta^i_m \delta^j_l) = 0 \;.
\end{equation}
The third transformation is $\delta Q_a^i=\epsilon^i_{\phantom{i}j}
Q_a^j M^n$, where $M \equiv (J^{-1})_{ij} M^{ij}$:
\begin{equation}
2 \lambda \ev{S_{ij} M^{lj} M^n} + \frac{2 N_c}{3} G^2 \ev{M^n}
\delta^l_i + \frac{2 n}{3} G^2 (J^{-1})_{ij} \ev{M^{lj} M^{n-1}} = S
\ev{M^n} \delta^l_i \;.
\end{equation}
Using $\delta S_{ij} = \epsilon_{ij}^{\phantom{ij}lm} S_{lm} \tilde S
^n$, where $\tilde S \equiv J^{ij} S_{ij}$ one obtains
\begin{eqnarray}
\lefteqn{\lambda \ev{S_{ij} M^{lm} \tilde S^n} - m J^{lm} \ev{S_{ij} \tilde
S^n} + 2 \alpha \ev{S_{ij} S^{lm} \tilde S^n}} \nonumber \\
&& {} + \frac{1}{6} G^2 \ev{\tilde S^n} (\delta^l_i \delta^m_j
- \delta^l_j \delta^m_i) +
\frac{n}{3} G^2 J^{lm} \ev{S_{ij} \tilde S^{n-1}} = 0 \;.
\end{eqnarray}
The mixing of fundamental and gauge-singlet matter $\delta Q_a^i =
\epsilon^i_{\phantom{i}l} Q_a^l \tilde S^n$ yields
\begin{equation}
2 \lambda \ev{S_{ij} M^{lj} \tilde S^n} + \frac{2 N_c}{3} G^2
\ev{\tilde S^n} \delta^l_i = S \ev{\tilde S^n} \delta^l_i \;.
\end{equation}
And with the transformation $\delta S_{ij} =
\epsilon_{ij}^{\phantom{ij}lm} S_{lm} M^n$ one obtains the equation
\begin{eqnarray}
\lambda \ev{S_{ij} M^{lm} M^n} - m J^{lm} \ev{S_{ij} M^n}
+ 2 \alpha \ev{S_{ij} S^{lm} M^n} \nonumber \\
{} + \frac{1}{6} G^2 \ev{M^n}
(\delta^l_i \delta^m_j - \delta^l_j \delta^m_i) =0 \;.
\end{eqnarray}
The transformation $\delta Q_a^i=\epsilon^i{}_j J^{kj} S_{kl} Q^l_a$
supplies us with the last equation we need
\begin{equation}
2 \lambda \ev{S_{kl} M^{lm} S_{im}} + \frac{2 N_c}{3} G^2 \ev{S_{ki}}
= S \ev{S_{ki}} \;.
\end{equation}

With the assumptions that connected three-point-functions vanish in
the chiral ring, that connected two-point functions are proportional
to $G^2$ and that the vacuum has flavor symmetry, the required
correlation functions can be parameterized as follows
\begin{eqnarray}
\ev{M^{lm}} & = & (M_0 + M_1 G^2) J^{lm} \;, \\
\ev{S_{ij}} & = & (S_0 + S_1 G^2) (J^{-1})_{ij} \;, \\
\ev{\tilde S} & = & -2 N_f (S_0 + S_1 G^2) \;,  \\
\ev{S_{ij} M^{lm}}_c & = & \left[ A (J^{-1})_{ij} J^{lm} + B
(\delta^l_i \delta^m_j - \delta^l_j \delta^m_i) \right] G^2 \;, \\
\ev{S_{ij} M}_c & = & 2(B - N_f A ) G^2 (J^{-1})_{ij} \;, \\
\ev{S_{ij} M^{lm}} & = & \left[ S_0 M_0 + (S_0 M_1 + S_1 M_0 +A) G^2
\right] (J^{-1})_{ij} J^{lm} \nonumber \\
&& {} + B G^2 (\delta^l_i \delta^m_j -
\delta^l_j \delta^m_i) \;, \\
\ev{S_{ij} S^{lm}}_c & = & \left[ C (J^{-1})_{ij} J^{lm} + D
(\delta^l_i \delta^m_j - \delta^l_j \delta^m_i) \right] G^2 \;, \\
\ev{S_{ij} S^{lm}} & = & \left[ -S_0^2 + (C - 2 S_0 S_1) G^2 \right]
(J^{-1})_{ij} J^{lm} \nonumber \\
&& {} + D G^2 (\delta^l_i \delta^m_j - \delta^m_i \delta^l_j) \;, \\
\ev{M^{ij} M^{lm}}_c & = & \left[E J^{ij} J^{lm} + F (J^{il} J^{jm} - J^{im}
J^{jl}) \right] G^2 \;, \\
\ev{M^{ij} M}_c & = & -2 (N_f E + F) G^2 J^{ij} \;, \\
\ev{M^{ij} M} & = & -2 \left[ N_f M_0^2 + (2 N_f M_0 M_1 + N_f E + F)
G^2 \right] J^{ij} \;, \\
\ev{S_{ij} M^{lj} M} & = & 2 \Bigg\{ N_f S_0 M_0^2 + \bigg[ N_f M_0
\Big( 2 S_0 M_1 + S_1 M_0 - (2 N_f - 1) B + 2 A \Big)
\nonumber \\
&& {} - M_0 B + S_0 ( N_f E + F) \bigg] G^2 \Bigg\} \delta^l_i \;, \\
\ev{S_{ij} \tilde S}_c & = & 2 (N_f C - D) G^2 (J^{-1})_{ij} \;, \\
\ev{S_{ij} \tilde S} & = & -2 \left\{ N_f S_0^2 + \left[ N_f \left(
2 S_0 S_1 - C \right) + D \right] G^2 \right\} (J^{-1})_{ij} \;, \\
\ev{M^{lm} \tilde S}_c & = & 2 (B- N_f A) G^2 J^{lm} \;, \\
\ev{S_{ij} M^{lm} \tilde S} & = & -2 \bigg\{ N_f S_0^2 M_0
+ \Big[ N_f S_0 (S_0 M_1 + 2 S_1 M_0 + 2 A) \nonumber \\
&& {} - M_0 (N_f C - D) - S_0 B \Big] G^2 \bigg\} (J^{-1})_{ij} J^{lm}
\nonumber \\
&& {} - 2 N_f S_0 B G^2 (\delta^l_i \delta^m_j - \delta^l_j
\delta^m_i) \;, \\
\ev{S_{ij} S^{lm} \tilde S} & = & 2 \left\{ N_f S_0^3
+ \left[ 3 N_f S_0 \left( S_0 S_1 - C \right) + 2 S_0 D \right] G^2
\right\} (J^{-1})_{ij} J^{lm} \nonumber \\
&& {} - 2 N_f S_0 D G^2 (\delta^l_i \delta^m_j - \delta^l_j
\delta^m_i) \;, \\
\ev{S_{ij} M^{lm} M} & = & -2 \Big\{ N_f S_0 M_0^2 + \big[ N_f M_0
(2 S_0 M_1 + S_1 M_0 + 2 A) -M_0 B \nonumber \\
&& {} + S_0 (N_f E + F) \big] G^2 \Big\} (J^{-1})_{ij} J^{lm}
\nonumber \\
&& {} - 2 N_f M_0 B G^2 (\delta^l_i \delta^m_j - \delta^l_j
\delta^m_i) \;, \\
\ev{S_{ij} S^{lm} M} & = & 2 \Big\{ N_f S_0^2 M_0 + \big[ N_f S_0
(S_0 M_1 + 2 S_1 M_0 + 2 A) -N_f M_0 C \nonumber \\ 
&& {} - 2 S_0 B \big] G^2 \Big\} (J^{-1})_{ij} J^{lm} -2 N_f M_0 D G^2
(\delta^l_i \delta^m_j - \delta^l_j \delta^m_i) \;, \\
\ev{S_{kl} M^{lm} S_{im}} & = & 2 S_0 [(2 N_f - 1) B - A] G^2
(J^{-1})_{ki} + M_0 [C - (2 N_f - 1) D] G^2 (J^{-1})_{ki}
\nonumber \\
&& {} + [S_0^2 M_0 + (S_0^2 M_1 + 2 S_0 S_1 M_0) G^2] (J^{-1})_{ik} \;.
\end{eqnarray}

The Konishi anomaly equations are then expressed as the ten equations
\begin{eqnarray}
&& 2 \lambda S_0 M_0 + S = 0 \;, \\
&& 2 \lambda \left[ (2 N_f -1 ) B - A -S_0 M_1 -S_1 M_0 \right]
+ \frac{2 N_c}{3} = 0 \;, \\
&& (\lambda M_0 -m -2 \alpha S_0) S_0 = 0 \;, \\
&& \lambda (S_0 M_1 + S_1 M_0 + A) - m S_1 + 2 \alpha (C - 2 S_0 S_1) =
0 \;, \\
&& \lambda B + 2 \alpha D + \frac{1}{6} = 0 \;, \\
\lefteqn{2 \lambda \Big[ N_f M_0 \big( 2 S_0 M_1 + S_1 M_0 - (2 N_f -
1) B + 2 A \big) - {}} \nonumber \\
&& {} - M_0 B + S_0 (N_f E + F) \Big] - \frac{2 N_f N_c +1}{3}
M_0 + N_f S M_1 = 0 \;, \\
\lefteqn{-\lambda \left[ N_f S_0 (S_0 M_1 + 2 S_1 M_0 + 2 A) -M_0 (N_f C - D)
- S_0 B \right] + {}} \nonumber \\
&& {} + m \left[ N_f (2S_0 S_1 - C) + D \right] + 2 \alpha \left[ 3 N_f S_0
(S_0 S_1 - C) + 2 S_0 D \right] \nonumber \\
&&  = -\frac{S_0}{6} \;, \\
\lefteqn{2 \lambda \big[ N_f S_0 (S_0 M_1 + 2 S_1 M_0 + 2 A) - M_0 (N_f C - D)
- {}}
\nonumber \\
&& {} - (N_f (2 N_f - 1) +1) S_0 B \big] - \frac{2 N_f N_c}{3} S_0
+ N_f S S_1 = 0 \;, \\
\lefteqn{-\lambda \left[ N_f M_0 ( 2 S_0 M_1 + S_1 M_0 + 2 A) - M_0 B
+ S_0 (N_f E + F) \right]} \nonumber \\
&& {} + m \left[ N_f (S_0 M_1 + S_1 M_0 + A) - B \right] \nonumber \\
&& {} + 2 \alpha \left[ N_f S_0 (S_0 M_1 + 2 S_1 M_0 + 2 A) -N_f M_0 C
- 2 S_0 B \right] = 0 \;, \\
\lefteqn{4 \lambda S_0 \left[ (2 N_f - 1) B -A \right] + 2 \lambda M_0 \left[
C - (2 N_f - 1) D \right]} \nonumber \\
&& {} - 2 \lambda (S_0^2 M_1 + 2 S_0 S_1 M_0) + \frac{2 N_c}{3} S_0 =
S S_1 \;,
\end{eqnarray}
whose solutions are
\begin{eqnarray}
S_0 & = & \frac{-m + \sqrt{m^2 - 4 \alpha S}}{4 \alpha} \;, \nonumber \\
S_1 & = & \frac{4 N_c \sqrt{m^2 - 4 \alpha s} - (2 N_f - 1) (m +
\sqrt{m^2 - 4 \alpha S})}{12 (m^2 - 4 \alpha S)} \;, \nonumber \\
M_0 & = & \frac{m + \sqrt{m^2 -4 \alpha S}}{2 \lambda} \;, \nonumber \\
M_1 & = & \frac{\alpha \left[ 4 N_c \sqrt{m^2 - 4 \alpha S} - (2 N_f - 1) (m
+ \sqrt{m^2 - 4 \alpha S}) \right]}{6 \lambda (m^2 - 4 \alpha S)} \;,
\nonumber \\
A & = & 0 \;, \\
B & = & - \frac{m^2 - 4 \alpha S + m \sqrt{m^2 - 4 \alpha S}}{12 \lambda
(m^2 -4 \alpha S)} \;, \nonumber \\
C & = & 0 \;, \nonumber \\
D & = & -\frac{m^2 - 4 \alpha S - m \sqrt{m^2 - 4 \alpha S}}{24 \alpha
(m^2 - 4 \alpha S)} \;, \nonumber \\
H & = & \frac{\alpha (m^2 - 4 \alpha S + m \sqrt{m^2 -4 \alpha S})}{6
\lambda^2 (m^2 - 4 \alpha S)} \;, \nonumber 
\end{eqnarray}
and
\begin{eqnarray}
S_0 & = & -\frac{m + \sqrt{m^2 - 4 \alpha S}}{4 \alpha} \;, \nonumber \\
S_1 & = & \frac{-4 N_c \sqrt{m^2 - 4 \alpha S} + (2 N_f - 1) (-m +
\sqrt{m^2 - 4 \alpha S})}{12 (m^2 -4 \alpha S)} \;, \nonumber \\
M_0 & = & \frac{m - \sqrt{m^2 -4 \alpha S}}{2 \lambda} \;, \nonumber \\
M_1 & = & \frac{\alpha \left[ -4 N_c \sqrt{m^2 -4 \alpha S} + (2 N_f
- 1) (-m + \sqrt{m^2 -4 \alpha S}) \right]}{6 \lambda (m^2 -4 \alpha
S)} \;, \nonumber \\
A & = & 0 \;, \label{eq:DSB-massive-vacuum-solution} \\
B & = & -\frac{m^2 - 4 \alpha S - m \sqrt{m^2 - 4 \alpha S}}{12
\lambda (m^2 -4 \alpha S)} \;, \nonumber \\
C & = & 0 \;, \nonumber \\
D & = & -\frac{m^2  -4 \alpha S + m \sqrt{m^2 -4 \alpha S}}{24 \alpha
(m^2 - 4 \alpha S)} \;, \nonumber \\
H & = & \frac{\alpha \left( m^2 - 4 \alpha S - m \sqrt{m^2 - 4 \alpha
S} \right)}{6 \lambda^2 (m^2 - 4 \alpha S)} \;, \nonumber
\end{eqnarray}
where $H \equiv N_f E+F$.
The solution (\ref{eq:DSB-massive-vacuum-solution}) corresponds to the
massive vacuum solution, in which the chiral multiplets are
massive.

\section{$G_2$ SYM with Three Flavors Details} \label{sec:G_2details}
Using the most simple Konishi transformation, which involves only a
flavor rotation $\delta Q^i_I=\lambda_I^{\phantom{I}J} Q^i_J$, we
obtain the Konishi equation
\begin{equation}
2 m \ev{X_{IJ}} + \lambda \ev{Z} \delta_{IJ} + \frac{7}{3} G^2
\delta_{IJ} = 2 S \delta_{IJ} \;,
\end{equation}
where the factor of two in front of the glueball superfield $S$ is due
to the representation index of the $\mathbf{7}$ representation of
$G_2$.  The transformation $\delta Q^i_I=\psi^{ijk} \epsilon_{IJK}
Q^j_J Q^k_K$ generates the equation
\begin{equation}
2 m \ev{Z} + 6 \lambda \left( \ev{X_I^{\phantom{I}I}
X_J^{\phantom{J}J}} - \ev{X_{IJ} X^{IJ}} \right) = 0 \;.
\end{equation}
A more general transformation is $\delta
Q^i_M=\lambda_M^{\phantom{M}L} Q^i_L \tr (X^n)$, where the trace is
taken on flavor indices. The resulting equation is
\begin{eqnarray}
\lefteqn{2 m \ev{X_{IL} \tr(X^n)} + \lambda \ev{Z \tr(X^n)} \delta_{IL} +
\frac{7}{3} G^2 \ev{\tr (X^n)} \delta_{IL}}
\nonumber \\
&& {} + \frac{2n}{3} G^2 \ev{(X^n)_{LI}} = 2 S \ev{\tr(X^n)}
\delta_{IL} \;.
\end{eqnarray}
A generalization of the second transformation, $\delta Q^i_I =
\psi^{ijk} \epsilon_{IJK} Q^j_J Q^k_K \tr (X^n)$, yields for
$n \ge 1$ the equation
\begin{eqnarray}
&& 2 m \ev{Z \tr (X^n)} + 6 \lambda \left(
\ev{X_I^{\phantom{I}I} X_J^{\phantom{J}J} \tr (X^n)} - \ev{X_{IJ}
X^{IJ} \tr (X^n)} \right) \nonumber \\
&& \quad {} + \frac{2 n}{9} G^2 \ev{Z \tr (X^{n-1})} = 0
\end{eqnarray}
The last required equation is obtained from
$\delta Q^i_I=\lambda_I^{\phantom{I}J} X_J^{\phantom{J}K} Q^i_K$:
\begin{equation}
2 m \ev{X_{JK} X^{KI}} + \lambda \ev{X_J^{\phantom{J}I} Z} +
\frac{1}{3} G^2 \ev{X_K^{\phantom{K}K}} \delta_J^{\phantom{J}I} +
\frac{8}{3} G^2 \ev{X_J^{\phantom{J}I}} = 2 S \ev{X_J^{\phantom{J}I}} \;.
\end{equation}

If we assume that the vacuum does not break the $SO(3)$ flavor
symmetry and that connected two-point functions are of order $G^2$
($\ev{\sigma_I \sigma_J}_c \sim G^2$), we may express the correlation
functions in the form
\begin{eqnarray}
\ev{X_{IJ}} & = & (A_0 + A_1 G^2) \delta_{IJ} \;, \\
\ev{Z} & = & Z_0 + Z_1 G^2 \;, \\
\ev{X_{IJ} X_{KL}}_c & = & (B \delta_{IJ} \delta_{KL} +
C \delta_{JK} \delta_{IL} + C \delta_{IK} \delta_{JL}) G^2 \;, \\
\ev{X_{IJ} X_K^{\phantom{K}K}}_c & = & ( 3 B + 2 C ) \delta_{IJ} G^2
\;, \\
\ev{Z X_{IJ}}_c & = & D \delta_{IJ} G^2 \;, \\
\ev{X_I^{\phantom{I}I} X_J^{\phantom{J}J}}_c & = & ( 9 B + 6 C ) G^2
\;, \\
\ev{X_I^{\phantom{I}I} X_J^{\phantom{J}J}} & = & 3 \left[ 3 A_0^2 +
\left( 3 B + 2 C + 6 A_0 A_1 \right) G^2 \right] \;, \\
\ev{X_{IJ} X^{JI}} & = & 3 \left[ A_0^2 + \left( B + 4 C + 2 A_0 A_1
\right) G^2 \right] \;, \\
\ev{X_{IJ} X_K^{\phantom{K}K}} & = & \left[ 3 A_0^2 + \left( 3 B + 2 C
+ 6 A_0 A_1 \right) G^2 \right] \delta_{IJ} \;, \\
\ev{Z X_I^{\phantom{I}I}} & = & 3 \left[ A_0 Z_0 + \left( D + A_0 Z_1
+ A_1 Z_0 \right) G^2 \right] \;, \\
\ev{X_{JK} X^{KI}} & = & \left[ A_0^2 + \left( 2 A_0 A_1 + B + 4 C
\right) G^2 \right] \delta_J^{\phantom{J}I} \;, \\
\ev{X_J^{\phantom{J}I} Z} & = & \left[ A_0 Z_0 + \left( A_0 Z_1 + A_1 Z_0 +D
\right) G^2 \right] \delta_J^{\phantom{J}I} \;.
\end{eqnarray}

By farther assuming that connected three-point functions vanish in the
chiral ring one gets
\begin{eqnarray}
\ev{X_I^{\phantom{I}I} X_J^{\phantom{J}J} X_K^{\phantom{K}K}} & = &
27 \left[ A_0^3 + A_0 \left( 3 B + 2 C + 3 A_0 A_1 \right) G^2 \right]
\;, \\
\ev{X_{IJ} X^{IJ} X_K^{\phantom{K}K}} & = & 3 \left[ 3 A_0^3 +
A_0 \left( 9 B + 16 C + 9 A_0 A_1 \right) G^2 \right] \;.
\end{eqnarray}

Using the Konishi equations up to $n=1$ yields the following equations
for the seven unknowns $A_0$, $A_1$, $Z_0$, $Z_1$, $B$, $C$ and $D$
\begin{eqnarray}
2 S & = & 2 m A_0 + \lambda Z_0 \;, \\
0 & = & 2 m A_1 + \lambda Z_1 + \frac{7}{3} \;, \\
0 & = & m Z_0 + 18 \lambda A_0^2 \;, \\
0 & = & m Z_1 + 18 \lambda \left( B - C + 2 A_0 A_1 \right) \;, \\
6 S A_1 & = & 2 m \left ( 3 B + 2 C + 6 A_0 A_1 \right) + 3 \lambda \left( D + A_0
Z_1 + A_1 Z_0 \right) + \frac{23}{3} A_0 \;, \\
0 & = & m \left( D + A_0 Z_1 + A_1 Z_0 \right) + 6 \lambda A_0 \left( 9 B + C +
9 A_0 A_1 \right) + \frac{1}{9} Z_0 \;, \\
2 S A_1 & = & 2 m \left( 2 A_0 A_1 + B + 4 C \right) + \lambda \left(
A_0 Z_1 + A_1 Z_0 + D \right) + \frac{11}{3} A_0 \;,
\end{eqnarray}
whose two solutions are
\begin{eqnarray}
A_0 & = & \frac{m^2 + m \sqrt{m^2 - 36 \lambda^2 S / m}}{18 \lambda^2}
\;, \nonumber \\
A_1 & = & \frac{-3 m^3 +  72 \lambda^2 S + 11 m^2 \sqrt{m^2 - 36
\lambda^2 S / m}}{12 m^2 \left( m^2 - 36 \lambda^2 S / m \right)} \;,
\nonumber \\
Z_0 & = & - \frac{m^3 - 18 \lambda^2 S + m^2 \sqrt{m^2 - 36 \lambda^2
S / m}}{9 \lambda^3} \;, \nonumber \\
Z_1 & = & - \frac{11 m^3 - 432 \lambda^2 S + 11 m^2 \sqrt{m^2 - 36
\lambda^2 S / m}}{6 \lambda m \left( m^2 - 36 \lambda^2 S / m\right)}
\;, \label{eq:G2solution1} \\
B & = & \frac{\left( m + \sqrt{m^2 - 36 \lambda^2 S / m}
\right)^2}{108 \lambda^2 \sqrt{m^2 -36 \lambda^2 S / m}} \;, \nonumber
\\
C & = & - \frac{m + \sqrt{m^2 - 36 \lambda^2 S / m}}{108 \lambda^2}
\;, \nonumber \\
D & = & - \frac{m \left( m^2 - 18 \lambda^2 S / m + m \sqrt{m^2 -36
\lambda^2 S / m}\right)}{27 \lambda^3 \sqrt{m^2 - 36 \lambda^2 S / m}}
\;, \nonumber
\end{eqnarray}
and
\begin{eqnarray}
A_0 & = & \frac{m^2 - m \sqrt{m^2 - 36 \lambda^2 S / m}}{18 \lambda^2}
\;, \nonumber \\
A_1 & = & - \frac{3 m^3 - 72 \lambda^2 S + 11 m^2 \sqrt{m^2 - 36
\lambda^2 S / m}}{12 m^2 \left( m^2 -36 \lambda^2 S / m \right)} \;,
\nonumber \\
Z_0 & = & \frac{- m^3 + 18 \lambda^2 S + m^2 \sqrt{m^2 -36 \lambda^2 S
/ m}}{9 \lambda^3} \;, \nonumber \\
Z_1 & = & - \frac{11 m^3 - 432 \lambda^2 S - 11 m^2 \sqrt{m^2 - 36
\lambda^2 S / m}}{6 \lambda m \left( m^2 - 36 \lambda^2 S / m \right)}
\;, \label{eq:G2solution2} \\
B & = & - \frac{\left( m - \sqrt{m^2 - 36 \lambda^2 S / m}
\right)^2}{108 \lambda^2 \sqrt{m^2 - 36 \lambda^2 S / m}} \;,
\nonumber \\
C & = & - \frac{m - \sqrt{m^2 - 36 \lambda^2 S / m}}{108 \lambda^2}
\;, \nonumber \\
D & = & \frac{m^3 - 18 \lambda^2 S - m^2 \sqrt{m^2 - 36 \lambda^2 S /
m}}{27 \lambda^3 \sqrt{m^2 -36 \lambda^2 S / m}} \;. \nonumber
\end{eqnarray}
Solution (\ref{eq:G2solution2}) corresponds to the Higgsed vacuum
found in \cite{Brandhuber:2003va}.

\section{$G_2$ Vector Model Diagrammatics} \label{sec:G2-diagrammatics}
The Feynman rules, read from the $G_2$ vector model action, are
\begin{center}
\input{G2rules.pstex_t}
\end{center}

The vector model free-energy is defined by
\begin{equation}
e^{-\mathcal{F}(g)} = Z \;,
\end{equation}
with $Z$ the partition function given in (\ref{eq:G2partition-fn}).
Using the rules and the identity
\begin{equation}
\psi^{ijk} \psi^{ilm} = \delta^{jl} \delta^{km} - \delta^{jm}
\delta^{kl} + (*\psi)^{jklm} \;,
\end{equation}
where $*\psi$ denotes the form dual to the 3-form $\psi$,
the diagrams required for its perturbative computation (depicted in
Fig.~(\ref{fig:G2diagrams})) can be computed:
\begin{center}
\begin{tabular}{cr}
$\frac{21}{2} \frac{1}{m} g$ & (a) \\
$\frac{189}{2} \frac{\lambda^2}{m^3} g$ & (b) \\
$\frac{15309}{4} \left( \frac{\lambda^2}{m^3} \right)^2 g^2$ & (c) \\
$-\frac{5103}{8} \left( \frac{\lambda^2}{m^3} \right)^2 g^2$ & (d)
\end{tabular}
\end{center}
and the free-energy is
\begin{equation} \label{eq:G2free-energy}
\mathcal{F}(g) = -\frac{21}{2} \frac{1}{m} g - \frac{189}{2}
\frac{\lambda^2}{m^3} g - \frac{25515}{8} \left(
\frac{\lambda^2}{m^3} \right)^2 g^2 + O(g^3) \;.
\end{equation}

If we \emph{assume} that we may obtain the perturbative effective
superpotential of the gauge theory by the identification
$g=\frac{2}{7}S$ from the relation (\ref{eq:G2-g-S-relation}), we get
under this assumption that the perturbative effective superpotential
should be
\begin{equation}
W_\mathrm{eff}^\mathrm{pert} \stackrel{?}{=} -\frac{2}{7} S^2
\frac{\partial \mathcal{F}(\frac{2}{7} S)}{\partial S} \;.
\end{equation}
This hypothesis is now tested using the perturbative result
(\ref{eq:G2free-energy}) and the exact effective superpotential
\cite{Brandhuber:2003va}. The diagrammatic result is
\begin{displaymath}
\frac{6}{7} \frac{1}{m} S^2 + \frac{54}{7} \frac{\lambda^2}{m^3} S^2 +
\frac{7290}{49} \left( \frac{\lambda^2}{m^3} \right)^2 S^3
\end{displaymath}
while the expansion of the exact result is 
\begin{equation}
W_\mathrm{eff}^\mathrm{pert} = (1 + \log 4 + 3 \log m) S - 9
\frac{\lambda^2}{m^3} S^2 -81 \left( \frac{\lambda^2}{m^3} \right)^2
S^3 +O(S^4) \;.
\end{equation}

The above suggests that that even the lowest order diagram
(Fig.~\ref{fig:G2diagrams}(b)) has a gravitational
contribution. Because of the structure of the vertex in this model
there is no obvious way to identify ``index loops'' and --- unlike
gauge groups that admit large-$N$ expansion --- it does not have an
expansion parameter that indicates the order of the gravitational
F-terms, and the non-gravitational contribution do not seem to be
diagrammatically isolated. Thus, it is conceivable, that this diagram
includes both $(\tr W_\alpha^2)^2 \sim S^2$ and $\tr[(W_\alpha^2)^2]
\sim S G^2$ terms with different coefficients.

\FIGURE[t]{
\includegraphics{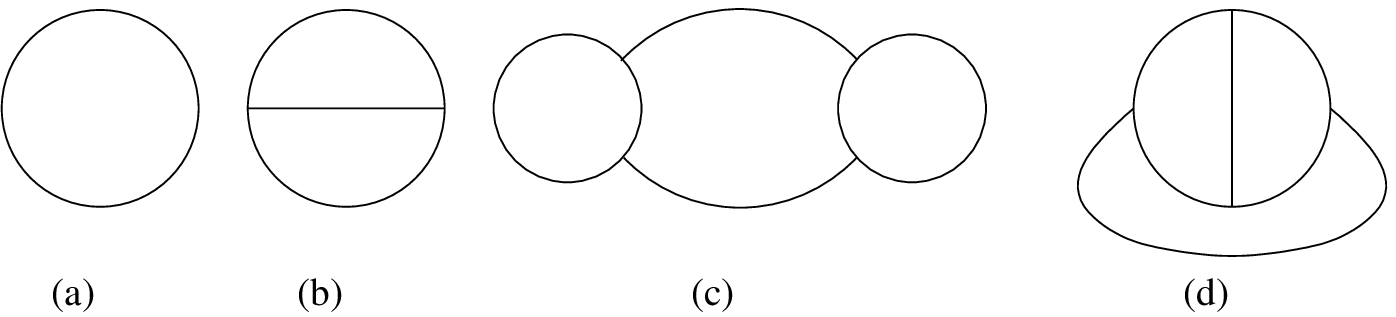}
\caption{The $G_2$ free-energy diagrams}
\label{fig:G2diagrams}
}

\newpage

\bibliographystyle{JHEP-2}
\bibliography{PhD}
\end{document}

%% file: G2rules.pstex_t
\begin{picture}(0,0)%
\includegraphics{G2rules.pstex}%
\end{picture}%
\setlength{\unitlength}{4144sp}%
\begingroup\makeatletter\ifx\SetFigFont\undefined
\def\x#1#2#3#4#5#6#7\relax{\def\x{#1#2#3#4#5#6}}%
\expandafter\x\fmtname xxxxxx\relax \def\y{splain}%
\ifx\x\y   
\gdef\SetFigFont#1#2#3{%
  \ifnum #1<17\tiny\else \ifnum #1<20\small\else
  \ifnum #1<24\normalsize\else \ifnum #1<29\large\else
  \ifnum #1<34\Large\else \ifnum #1<41\LARGE\else
     \huge\fi\fi\fi\fi\fi\fi
  \csname #3\endcsname}%
\else
\gdef\SetFigFont#1#2#3{\begingroup
  \count@#1\relax \ifnum 25<\count@\count@25\fi
  \def\x{\endgroup\@setsize\SetFigFont{#2pt}}%
  \expandafter\x
    \csname \romannumeral\the\count@ pt\expandafter\endcsname
    \csname @\romannumeral\the\count@ pt\endcsname
  \csname #3\endcsname}%
\fi
\fi\endgroup
\begin{picture}(2610,1470)(1,-745)
\put(1261,569){\makebox(0,0)[lb]{\smash{\SetFigFont{12}{14.4}{rm}$j$,$J$}}}
\put(181,569){\makebox(0,0)[lb]{\smash{\SetFigFont{12}{14.4}{rm}$i$,$I$}}}
\put(1711,209){\makebox(0,0)[lb]{\smash{\SetFigFont{12}{14.4}{rm}$j$,$J$}}}
\put(  1,-241){\makebox(0,0)[lb]{\smash{\SetFigFont{12}{14.4}{rm}$i$,$I$}}}
\put(1711,-691){\makebox(0,0)[lb]{\smash{\SetFigFont{12}{14.4}{rm}$k$,$K$}}}
\put(2611,-241){\makebox(0,0)[lb]{\smash{\SetFigFont{12}{14.4}{rm}$-\frac{1}{g} \lambda \psi^{ijk} \epsilon_{IJK}$}}}
\put(2611,569){\makebox(0,0)[lb]{\smash{\SetFigFont{12}{14.4}{rm}$\frac{1}{2m} g \delta^{ij} \delta_{IJ}$}}}
\end{picture}